# Transmon qubit with relaxation time exceeding 0.5 milliseconds

Chenlu Wang, Xuegang Li, Huikai Xu, Zhiyuan Li, Junhua Wang, Zhen Yang, Zhenyu Mi, Xuehui Liang, Tang Su, Chuhong Yang, Guangyue Wang, Wenyan Wang, Yongchao Li, Mo Chen, Chengyao Li, Kehuan Linghu, Jiaxiu Han, Yingshan Zhang, Yulong Feng, Yu Song, Teng Ma, Jingning Zhang, Ruixia Wang, Peng Zhao, Weiyang Liu, Guangming Xue*, Yirong Jin**, Haifeng Yu***

Beijing academy of quantum information sciences, Beijing, China, 100193

**Abstract:**

By using the dry etching process of tantalum (Ta) film, we had obtained transmon qubit with the best lifetime ($T_1$) 503 μs, suggesting that the dry etching process can be adopted in the following multi-qubit fabrication with Ta film. We also compared the relaxation and coherence times of transmons made with different materials (Ta, Nb and Al) with the same design and fabrication processes of Josephson junction, we found that samples prepared with Ta film had the best performance, followed by those with Al film and Nb film. We inferred that the reason for this difference was due to the different loss of oxide materials located at the metal-air interface.

**Introduction:**

Superconducting quantum computing (SQC) has developed rapidly in recent years. Thanks to its compatibility with semiconductor technology, now it has entered the stage of scaling. Quantum supremacy has been demonstrated in a 53-bit SQC system[1], showing SQC has a promising future in realizing a practical quantum computer. Superconducting qubit is an artificial system with large size (usually 300-500 micrometers). Compared with other natural candidates of quantum computer, such as ion traps, cold atoms and NV centers, the coherence time of superconducting qubit is shorter. This may limit its further development since quantum coherence time is the basis of gate manipulation and state evolution. After more than 20 years of development, with continuous efforts of many SQC groups, the coherence time of superconducting qubits has leaped from the initial order of nanoseconds to the order of hundreds of microseconds[2]. Although this is a remarkable progress, it still cannot fully meet the demand of the development of SQC, especially for quantum error correction[3] . As we know, the fidelity of the gate is closely related to the coherence time and higher gate fidelity means that the depth of the circuit can be deeper. For large-scale multi-qubit systems of next generation, the need for long coherence time is more urgent. In other words, the exploration of qubit with long coherence time is a never-satisfying and never-ending process.

Transmon[4] or Xmon[5] is the most used qubit in SQC. The principles of transmon and Xmon are the same except that one capacitor pad of Xmon is ground. Compared with other superconducting qubits, such as fluxonium[6] and C-shunt flux qubit[7,8], transmon qubit has a simple structure. The transmon qubit consists of two parts, one is the Josephson junction, which is usually made of aluminum material by the double-angle electron beam evaporation method. Up to now, no junction preparation that is easier and better than this method in SQC has been found to the best of our knowledge. This part of the fabrication process is relatively standard. The other part of transmon is the capacitor pad and other external circuits (such as readout cavity, transmission line, microwave drive and flux bias line, etc.). Elemental materials of Nb[9,10], Al[3,11], Ta[12] and compound materials of TiN[13], NbN[14], NbTiN[15], and ganular aluminum[16] are currently used. Elemental materials are more popular because film depositions are relatively simple and stable. In 2020, Princeton group made a breakthrough in the exploration of coherence time of transmon using Ta material[12]. Both the relaxation and coherence times exceed 0.3 milliseconds, showing an important improvement of coherence time of transmon. In their etching process, the quality of the wet-etched samples is better than that of dry-etched ones. For the development of large-scale quantum processors, the exploration of dry etch is necessary. In this paper, we develop and optimize dry etching process of Ta film and prepare samples. Results show that transmons with long coherence time can be obtained using dry etching process. In addition, we also compare the relaxation and coherence times of Ta transmons with Nb and Al transmons. All samples are prepared without changing fabrication processes except for the materials of qubit pads (Nb, Al and Ta). Experimental data show that without surface treatment, tantalum oxide has the lowest loss, followed by aluminum oxide and niobium oxide. These results suggest that Ta is a promising candidate for preparing superconducting qubits and can be used for further scalable quantum chip.

**Sample design and fabrication:**

In order to obtain a longer coherence time, we have optimized the sample design as follows. First, a single-junction design of fixed-frequency was adopted, which greatly reduces the influence of flux noise. Second, to ensure a quiet measurement environment, the chip had only two bonding pads of feedline, deep attenuation and filtering were careful carried out in the microwave semi-rigid cables, and no other control lines were introduced. In order to obtain statistical data, we placed five transmons on the chip and used the same feedline to drive and read them. Finally, in the design of transmon, we referred to the large pad adopted by Princeton and IBM groups to reduce the impact of surface participation[9]. The optical microscope photo of one sample is shown in Figure 1.

By using the above design, Nb, Al, and Ta films were used to prepare transmon samples. The preparation process is as follows. A superconducting film with a thickness of 120 nm was deposited on a chemically cleaned and annealed sapphire substrate. Here, the Nb and Ta films were prepared by magnetron sputter, and the Al film was prepared by electron beam evaporator. Transmon pads and readout resonators were patterned by ultraviolet (UV) lithography. We used inductively coupled plasma (ICP) or reactive ion etching (RIE) etcher to etch the superconducting films. After etching and the removal of UV photoresist, Dolan bridges were prepared by electron beam lithography (EBL) using PMMA A4/LOR10B double-layer photoresist. Al Josephson junctions were prepared using four-chamber E-beam evaporator (AdNaNoTek JEB4). Before deposition of Al film, an ion source with a radio frequency was used to clean the oxide layer on the Nb (Al or Ta) surface to achieve superconductivity connection. After wafer dicing, a liftoff process in N-Methylpyrrolidone (NMP) solution was used to remove photoresist. Finally, the chip was wire-bonded into a copper sample box. In order to obtain the best film quality, we optimized the deposition conditions of Nb and Ta films, especially for Ta film in alpha phase, including sputtering pressure, deposition speed, working distance between target and substrate, substrate temperature, etc. Finally, we obtained Nb film with a residual resistance ratio (RRR) 300K/10K of

4.9 and a critical temperature ($T_c$) of 9.1 K, while for Ta film RRR =4.5 and $T_c$=4.2 K. Figure 2 shows the X-ray diffraction (XRD) result of a Ta film in alpha phase.

In exploring the dry etching process of Ta, we used two etching machines, one is ICP etcher (PlasmaPro 100 Cobra, Oxford instrument), which has two radio frequency sources, and the other is RIE etcher (200NL, Samco). After multiple rounds of process optimization, we finally obtained the optimized etching parameters: ICP etcher ($SF_6$:$CHF_3$=4:1, 4 mTorr, powers of ICP and bias are 220 Watts and 50 Watts, respectively, the etching time is 180 seconds for Ta film with a thickness of 150 nm), RIE etcher ($CF_4$, 15 mTorr, 100 Watts and 180 seconds for Ta film with the same thickness). Figure 3 shows the SEM photos of different angles prepared by ICP etcher and RIE etcher. From the SEM photos, we could conclude that the etching effects of the two etchers are the same, and both can be used to prepare transmon. The Ta transmons in this paper were prepared by the RIE etcher with $CF_4$ gas.

**Measurement results and analysis**

We have prepared eight chips of different batches and put them into a low-noise dilution refrigerator for measurement [Supplementary materials]. The specific coherent properties are listed in Table 1.

Table 1：Parameters of different qubits of eight chips

| # Chips | Frequencies of $Q_1$-$Q_5$ (GHz) | $T_1$ (μs) of $Q_1$-$Q_5$ | $T_2^*$ (μs) of $Q_1$-$Q_5$ | $T_2^e$(μs) of $Q_1$-$Q_5$ |
|---|---|---|---|---|
| Nb-1 | 5.253/ 5.203/ 5.214/ 5.202 /5.515 | 26.5 /24.9 /21.6/ 26.5/ 19.2 | 4.6 /14.1/ 6.6/ 13.8 /0.8 | N/A |
| Nb-2[1] | 3.91/3.894/ 3.894 | 29.8/ 29.2/ 33.1 | 0.8 /10.5/ 9.5 | N/A |
| Al-1 | 4.562 /4.542/ 4.523 / 4.546 /4.549 | 92.6/78.8/ 21.2 / 58.3/ 43.6 | 2.2 /1.8 /2.7 / 3.0 /4.7 | N/A |
| Al-2[1] | 4.364 /4.339 /4.181/ 4.299 | 35.3/ 25.8 /107.8 / 63.6 | 10.8/ 12.4/ 11.3/ 10.9 | N/A |
| Ta-1 | 4.450/ 4.559/ 4.412/ 4.418 /4.502 | 158.3/ 109.2/ 136.3/ 120.9/ 131.2 | 75.5 /62.3/ 42/ 52.1/ 65.2 | 194.2/ 131.3/ 176/ 173.8/ 168.9 |
| Ta-2[3] | 4.688/ 4.659/ 4.608/ 4.681/ 4.735 | 359/ 158/ 102.7 / 347 /158 | 45 /117 /161/ 305/ 207 | 386 /317/ 205/ 431 /316 |
| Ta-3[1][3] | 4.294/4.304/ 4.264/ 4.422 | 341/225/372/ 337 | N/A[2] | 412/306/336/ 182 |
| Ta-4 | 3.894/3.913/3.918 3.864/3.890 | 329.1/476/401/ 312.8/316.8 | N/A[2] | 400.3/462.9/153.7/ 404/402.5 |

（1）Two transmons in Nb-2 chip were not measured. Same for one transmon in Al-2 chip and one transmon in Ta-3.

（2）$T_2^*$ values were measured and fitted using Ramsey experiment. We found that transmons on Ta-3 and Ta-4 chips had beating in their Ramsey oscillations and those will make the fitting value (T2*) not accurate, so we did not fill them in this table. $T_2^e$ values were measured and fitted

*using spin echo technology with one π pulse inserted between two π/2 pulses.*

*（3）The fabrication processes of chips Ta-1, Ta-2 ,Ta-3 and Ta-4 are the same except Ta-1 was not dipped into piranha solution before coating EBL photoresists.*

It can be seen from the table that the performance of the transmons with Ta films was better than those of Al and Nb films. $Q_3$ of sample Ta-4 was measured repeatedly.The average of $T_1$ was 401 μs and the best one reached 503 μs. Similarly, for $Q_1$ of Ta-3 with an averaged $T_1$ equal to 356 μs, the best one was 383 μs, $Q_1$ of Ta-2 with an averaged $T_1$ equal to 359 μs, the best one was 431 μs (Figure 4) . We haven't repeated $T_2^*$ and $T_2^e$ experiments over a long time. Carr-Purcell-Meiboom-Gill (CPMG) experiment [17] was also carried out in $Q_1$ of sample Ta-2. Here, two π pulses were inserted between two π/2 pulses and a value of 557 μs (Figure 5) was obtained by the fitting. We haven't found the increasing of this value with the number of inserted pulses over two.

We analyze the reasons of different relaxation times of different materials. As we know, the loss of transmon mainly comes from the interface. There are three types of interface related to metal films, which are metal-substrate (MS) interface between Nb (Al or Ta) film and sapphire substrate, metal-metal (MM) interface between Nb (Al or Ta) and Al Junction film and metal-air (MA) interface between Nb (Al or Ta) and air. The sapphire substrates we used here are carefully handled by chemical cleaning and annealing before film deposition. We heated them in a load-lock chamber (200 ℃ , 2 hours) for degassing. We believe that the MS interface has little effect on the difference in coherence performance of these samples. The same was true for MM interface; before fabrication of Josephson junction, Nb (Al or Ta) was cleaned with a radio frequency ion source and over-etched for 30 seconds to ensure that the oxide layer was removed completely. Although the interface properties between Nb-Al, Al-Al and Ta-Al were different, there were no contaminants introduced in this part during fabrication process. Therefore, we believe that the MA interface was the main factor causing this difference of coherence, because the metal was exposed to the air, and various components in the air can form compounds (mainly oxides) with the metal. These oxides contain various defects that will couple with the qubit and cause decoherence. The defects of this kind of oxide have been carefully studied by many research groups. Niobium oxides have three components: NbO, $NbO_2$, and $Nb_2O_5$ [18,19]. Verjauw et al. [20] found that if the niobium oxides were removed by hydrofluoric (HF) acid etching, the internal quality factor of resonator could reach up to 7 millions at single-photon power level, while that of the reference sample with intact native oxides is 1 million. This provides solid evidence that the two-level-system (TLS) defects located at the MA interface play an important role in affecting the decoherence of Nb transmon. Tantalum oxide has only one component, named $Ta_2O_5$. As shown in Figure 6, we have measured 3 samples (one is cleaned in piranha solution) using X-ray photospectroscopy (XPS) to obtain chemical elements of Ta film surface. The two peaks at low binding energy belong to $4f^{7/2}$ and $4f^{5/2}$ orbitals of tantalum metal, and the two peaks at higher binding energy correspond to the same orbitals of $Ta_2O_5$ [8]. We do not see any other chemical components except for $Ta_2O_5$, which means that $Ta_2O_5$ was the only source of defects on the surface of Ta film. From our results, it can be seen that without surface treatment, the results of Ta films were better than those of Nb films and Al films. However, if the surface oxide layers of the Ta films and the Nb films have been cleaned (because the cleaning process was not selective, the oxide layer on the surface of Al cannot be removed without damaging the Josephson junction), for example, by using HF vapor to remove oxide and using vacuum packaging method, it is difficult to determine which material has good performance, and further experiments are needed to verify.

We have made a comparison with a sample specially design with high MA SPR, as shown in Figure 7. By using the flip-chip technology, we add a patterned Ta film over the Ta transmon with a height of 5 micrometers (the total capacitance of this type of qubit is equal to the table 1 used, with the charging energy $E_c/2\pi$ ~ 260 MHz). The purpose of this design is to increase the SPR of MA deliberately . $T_1$ and $T_2^*$ are 50 μs and 75 μs measured, respectively, which are much lower than the conventional design of Figure 1. The shrinked footprint has not caused a sharp decrease of $T_1$. A more detailed data analysis will be presented in reference [21].

Practical superconducting quantum computing requires large-scale integration of qubits. In order to verify the effect of multiple qubits on quantum coherence, we prepared multi-qubit samples using the same Ta-transmon fabrication process above. The purcell limit of qubit design exceeds 1 ms. Figure 8 shows $T_1$, $T_2^*$, and $T_2^e$ of each fixed-frequency qubit in a chip of 56 qubits and 55 couplers ( $T_1$, $T_2^*$, and $T_2^e$ of each couplers were not plotted), all of which are much lower than those listed in Table 1. We partly attribute it to the environmental noise of the chip. For single qubit, the semi-rigid cables of control and measurement have deep attenuation and filtering, however, the same situation cannot be achieved in a multi-qubit chip for quick qubit control and flux bias. This will cause high-frequency noise to couple to the sample through cables and result in decoherence. Therefore, the environmental noise of the measurement system should be decreased in order to improve the decoherence time of large-scale samples. This is a challenging project at the current stage. To obtain long-lived qubits in a multi-qubit chip, it is significant to improve measurement setup as well as sample fabrication.

**Summary and prospects:**
In this article, we use the dry etching process of Ta film to prepare some single-qubit samples with the best lifetime of more than 500 μs. These results showed that dry etching can be adopted in the subsequent preparation of tantalum qubits. We also use the same design to compare the coherence time of qubits prepared with Nb, Al, and Ta, and find that the performance of Ta is the best. We believe that the loss of tantalum oxide is less than that of niobium oxide and aluminum oxide. In addition, we deliberately design a flip-chip transmon with a high SPR of MA. The comparison of $T_1$ and $T_2^e$ between flip-chip transmon and conventional one shows that dielectric loss from interfaces was still the main factor affecting the decoherence of qubits. We hope that these results are helpful for the preparation of long-lived qubits. Although one of the shortcomings of superconducting qubit is its short lifetime, although the improvement and optimization of the material interface, we believe that a coherence time of milliseconds could be achieved.

Acknowledgement: We thank Dr. Rui Wu for her fruitful suggestions on Ta film growth. This work was supported by the NSF of Beijing (Grants No. Z190012), the NSFC of China (Grants No.

11890704, 12004042), National Key Research and Development Program of China (Grant No. 2016YFA0301800) and the Key-Area Research and Development Program of Guang Dong Province (Grants No. 2018B030326001).

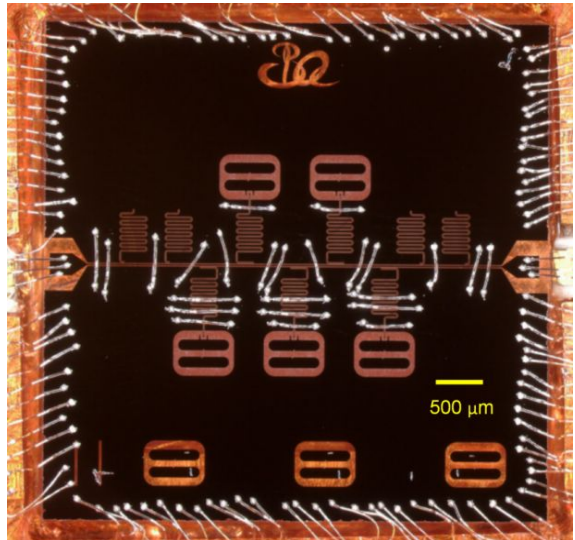

*Figure 1: An optical image of a packaged transmon sample with Ta film. The size of the chip is 7 mm. It contains five independent transmons, four independent quarter-wave resonators for measuring intrinsic Q factor of the resonator, and three transmons for resistance test. Purcell limit of transmon design is over 2 milliseconds. The coupling strength between readout resonator and transmon is 50 MHz ✕2π (experimental values are close to the design values).*

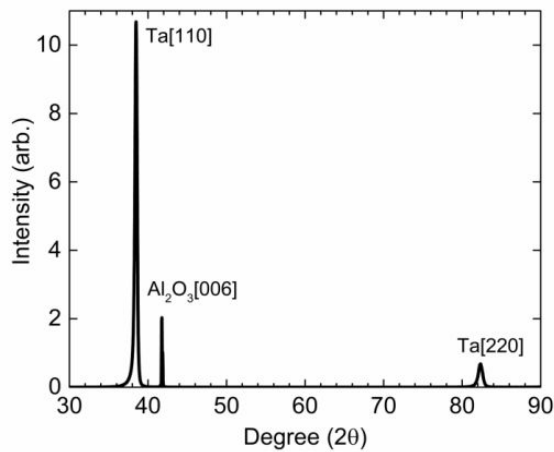

*Figure 2: XRD spectrum of a Ta film on sapphire. Two clear peaks correspond to [110] and [220] of alpha phase of Ta, and the peak in the middle corresponds to the sapphire substrate we used.*

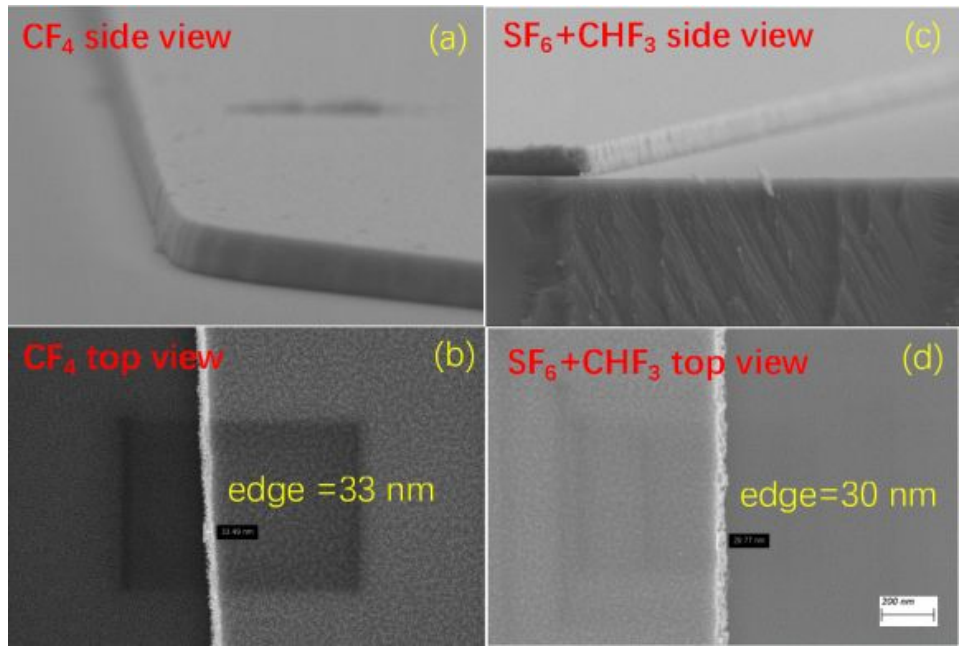

*Figure 3: Four SEM images of Ta film after dry etching. (a) and (b) are different angles of the same sample, which was etched in RIE etcher using $CF_4$ gas. Same for (c) and (d), but the $SF_6$:$CHF_3$ mixed gases were used in ICP etcher. Measured widths of the film edge of both samples were around 30 nm.*

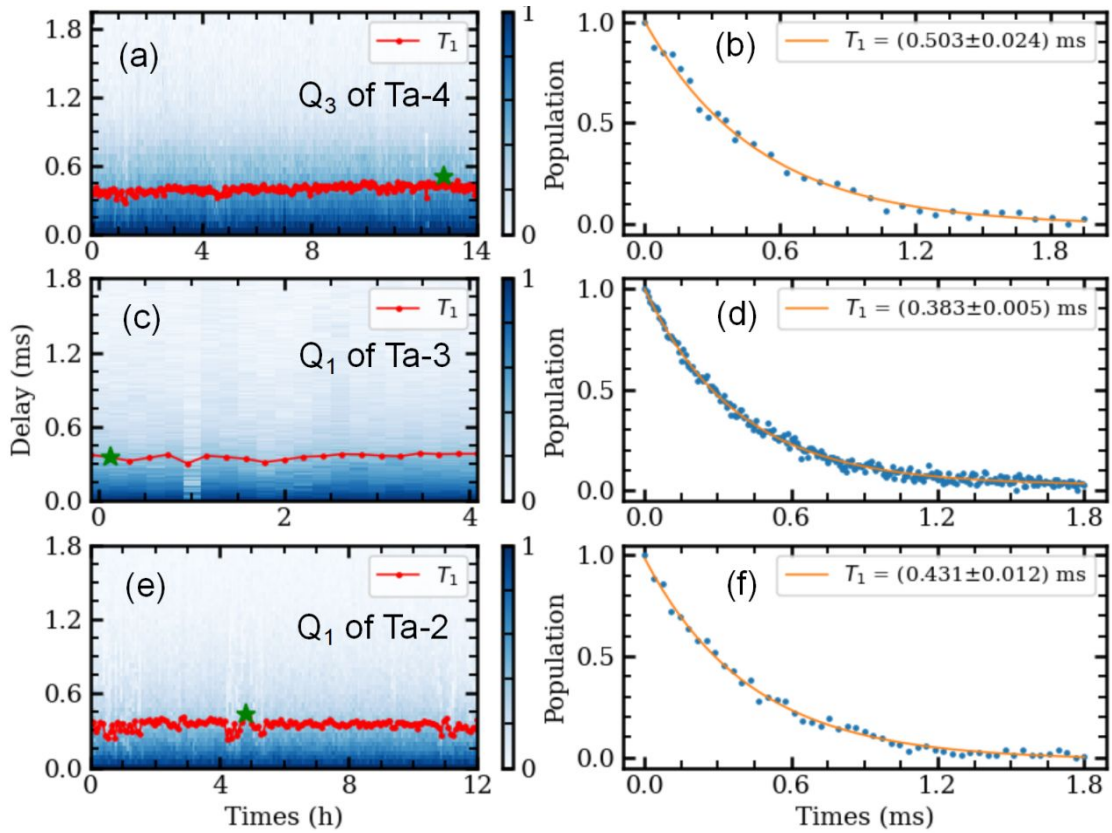

*Figure 4 : The relaxation time of $Q_3$ in Ta-4 chip (a), $Q_1$ in Ta-3 chip (c), $Q_1$ in Ta-2 chip (e) measured with 14 hours, 4 hours and 12 hours respectively. The red dot-lines are the fitting values of each decay curve; the average and best time are 401 µs and 503 µs (b) of $Q_3$ in Ta-4, sames are 356 µs and 383 µs (d) of $Q_1$ in Ta-3, 359 µs and 431 µs (f) of $Q_1$ in Ta-2.*

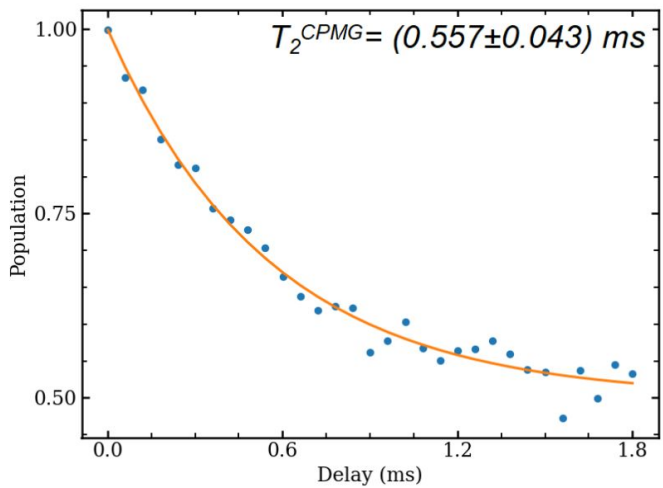

*Figure 5 : Coherence time of $Q_1$ in the Ta-2 chip measured using CPMG technology. $T_2^{CPMG}$ of 557 µs is obtained by the fitting.*

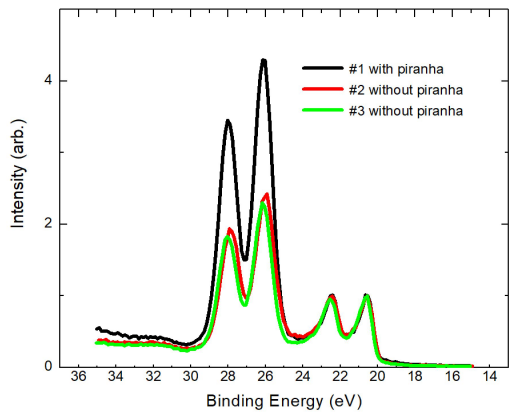

*Figure 6: The XPS spectra of three Ta samples. Two peaks at low binding energy belong to $4f^{7/2}$ and $4f^{5/2}$ orbitals of tantalum metal, and two peaks at higher energy correspond to the same orbitals of $Ta_2O_5$. The black line indicates that the sample has been dipped into piranha solution for 20 minutes with a temperature over 70 ℃, and shows thicker tantalum oxide than others that are not dipped into piranha solution.*

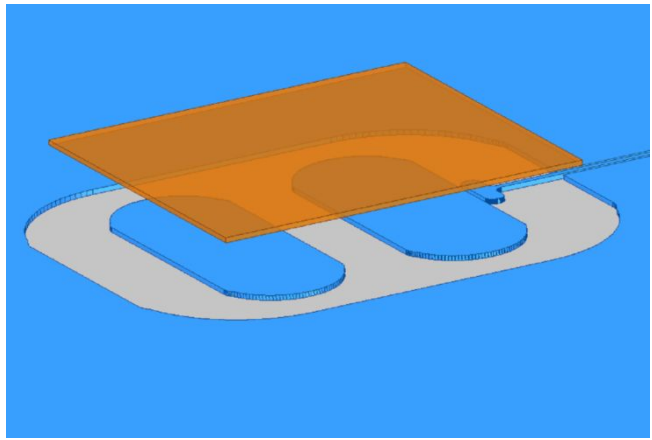

*Figure 7: A schematic picture of a transmon prepared by flip-chip process. It consists of two substrates with patterned tantalum films. The two substrates are packaged face-to-face with indium bumps using the flip-chip technology. The Josephson junction was located between the two pads (blue color) on the bottom substrate (Josephson junction was not shown in this picture), and the Ta film pattern (orange color) on the upper substrate was located directly above the two pads of the bottom substrate. The purpose of this is to increase the SPR of the MA interface, because there is an electric field that is localized between the upper and lower substrates.*

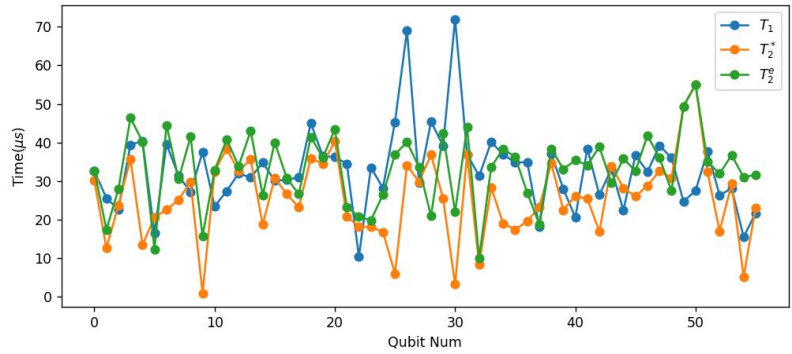

Figure 8: $T_1$, $T_2^*$, and $T_2^e$ of each qubit of a multi-qubit chip.

Corresponding author:
* xuegm@baqis.ac.cn
** jinyr@baqis.ac.cn
*** hfyu@baqis.ac.cn

**Supplementary materials of " Transmon qubit with relaxation time exceeding 0.5 milliseconds"**

Chenlu Wang, Xuegang Li, Huikai Xu, Zhiyuan Li, Junhua Wang, Zhen Yang, Zhenyu Mi, Xuehui Liang, Tang Su, Chuhong Yang, Guangyue Wang, Wenyan Wang, Yongchao Li, Mo Chen, Chengyao Li, Kehuan Linghu, Jiaxiu Han, Yingshan Zhang, Yulong Feng, Yu Song, Teng Ma, Jingning Zhang, Ruixia Wang, Peng Zhao, Weiyang Liu, Guangming Xue*, Yirong Jin**, Haifeng Yu***

Beijing academy of quantum information sciences, Beijing, China, 100193

**Measurement setup:**

The measurement circuit schematic is shown in Figure S1. The superconducting qubit was mounted in a copper sample box, as shown in Figure S2. It was anchored on a copper cold finger, and the cold finger has close contact with the 10mK mixing chamber (MC) plate of the dilution refrigerator to ensure good thermal contact. Eccosorb absorber and Cryoperm shielding were used to protect the chip from infrared radiation and flux noise. The microwave pulse used to drive qubit and readout cavity passed through a total 78dB attenuator (see Figure S1 for the attenuator installation of each cold plate) , a low-pass filter (LPF) with cut-off frequency of 8GHz and an infrared (IR) filter before reaching the chip. The output signal passed through an IR filter, a LPF and three cryo-circulators on the MC plate. After amplified by the High Electron Mobility Transistor (HEMT) on the 3K plate and amplifier at room temperature, the output signal reached Inphase Quadrature (IQ) mixer. The IQ mixer was used to convert microwave frequency signal to Intermediate frequency (IF) signal, the latter was amplified and then sent to the analog-to-digital conversion(ADC) card. We used direct digital synthesizers(Tektronix 70000 with sample rate 25GSa/s with each channel) to drive qubits and readout resonators. Compared with the method of using an IQ mixer, the interference of LO leakage and IQ imbalance can be avoided. Figure S3 is an optical photo of cryogenic circulator (4GHz-8GHz), LPF (cut-off frequency of 8GHz) and IR filter we used in this experiment.

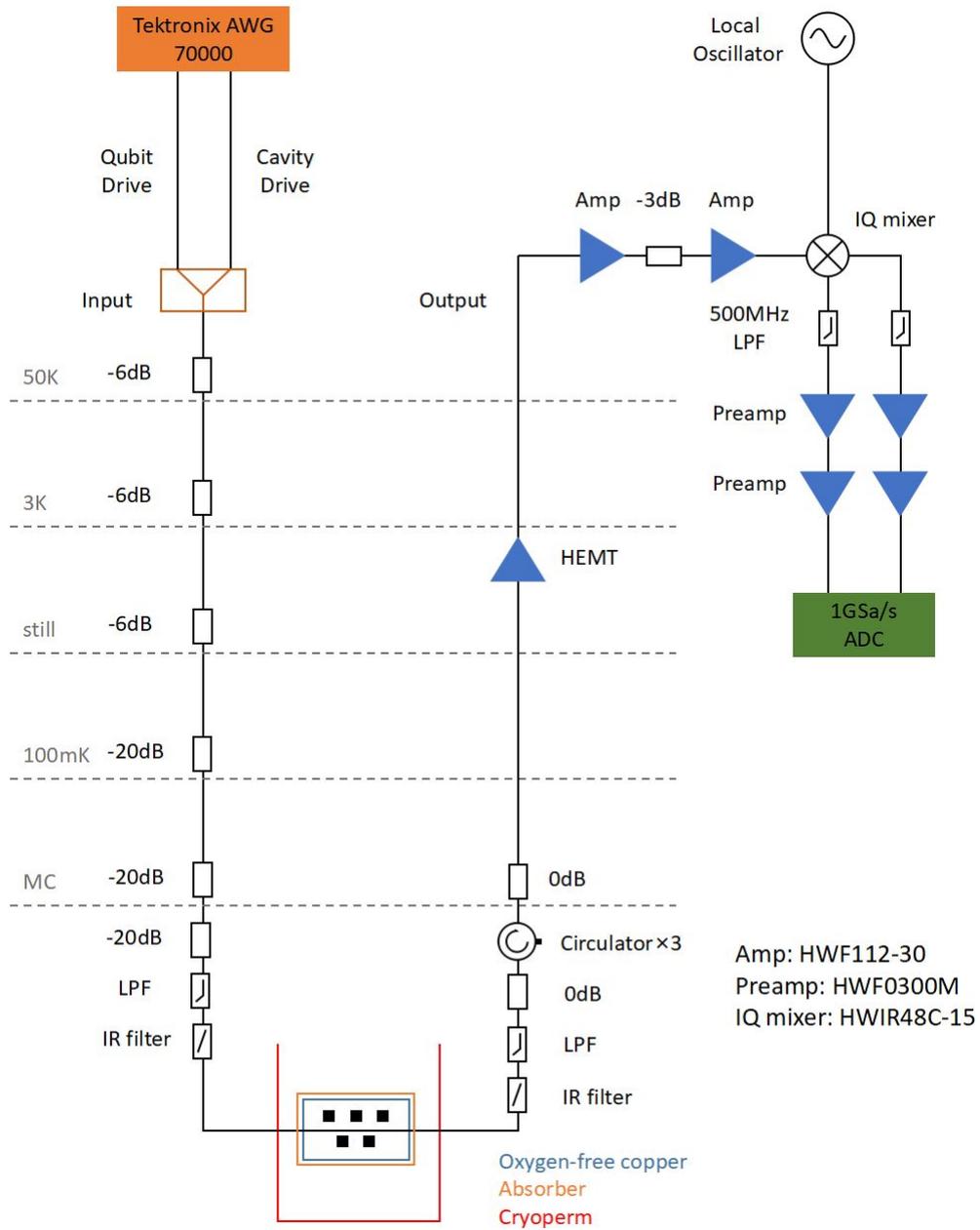

Figure S1, The schematics of measurement setup.

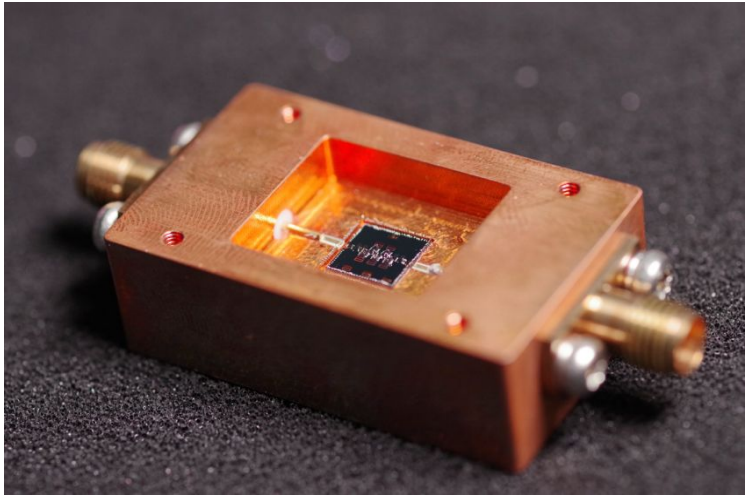

Figure S2, a photo of copper sample box used in our experiment (lid not shown).

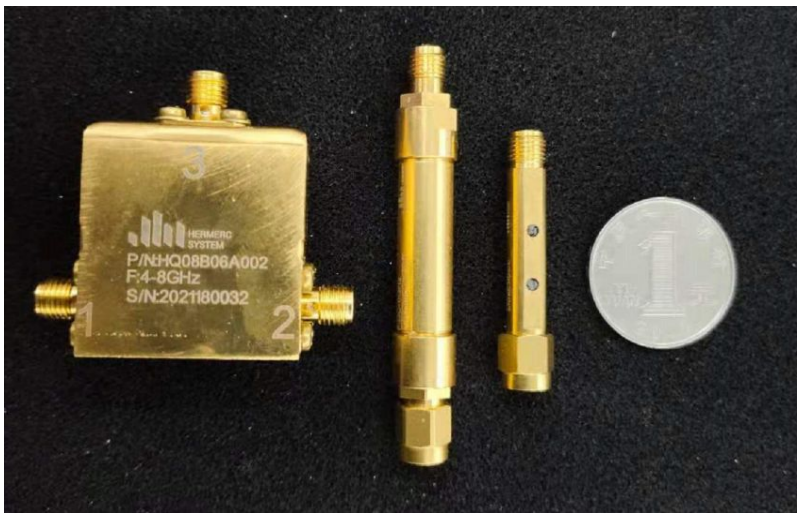

Figure S3, a photo of a cryo-circulator, a LP filter and an IR filter (from left to right ), they are all from Hermetic Inc , China. A coin of one Chinese Yuan is used as a reference of size.